\documentclass[twocolumn,showpacs,preprintnumbers,amsmath,amssymb]{revtex4}

\usepackage{graphicx}
\usepackage{bm}
\usepackage{amsmath}
\usepackage{color}

\begin{document}

\title{Ultrafast filling of an electronic pseudogap in an incommensurate crystal}

\author{V. Brouet$^1$, J. Mauchain$^1$, E. Papalazarou$^1$, J. Faure$^{2,3}$, M. Marsi$^1$, P.H. Lin$^1$, A. Taleb-Ibrahimi$^4$, P. Le F\`{e}vre$^4$, F. Bertran$^4$, L. Cario$^5$, E. Janod$^5$, B. Corraze$^5$, V. Ta Phuoc$^6$, and L. Perfetti$^2$}

\affiliation{$^{1}$ Laboratoire de Physique des Solides, CNRS-UMR 8502, Universit\'e Paris-Sud, F-91405 Orsay, France}
\affiliation{$^{2}$ Laboratoire des Solides Irradi\'{e}s, Ecole polytechnique-CEA/DSM-CNRS
UMR 7642, F-91128 Palaiseau, France}
\affiliation{$^{3}$ Laboratoire d'Optique Appliquée, UMR 7639 CNRS-ENSTA-Ecole Polytechnique, Palaiseau, France}
\affiliation{$^{4}$ Synchrotron SOLEIL, Saint-Aubin-BP 48, F-91192 Gif sur Yvette, France}
\affiliation{$^{5}$ Institut des mat\'eriaux Jean Rouxel, CNRS-UMR6502, Universit\'e de nantes, 2 rue de la Houssinière,44322 Nantes, France }
\affiliation{$^{6}$ GREMAN, CNRS UMR 7347 - CEA, Universit\'e F. Rabelais, UFR Sciences,
Parc de Grandmont, 37200 Tours, France}

\begin{abstract}

We investigate the quasiperiodic crystal (LaS)$_{1.196}$VS$_2$ by angle and time resolved photoemission spectroscopy. The dispersion of electronic states is in qualitative agreement with band structure calculated for the VS$_2$ slab without the incommensurate distortion. Nonetheless, the spectra display a temperature dependent pseudogap instead of quasiparticles crossing. The sudden photoexcitation at 50 K induces a partial filling of the electronic pseudogap within less than 80 fs. The electronic energy flows into the lattice modes on a comparable timescale. We attribute this surprisingly short timescale to a very strong electron-phonon coupling to the incommensurate distortion. This result sheds light on the electronic localization arising in aperiodic structures and quasicrystals.

\end{abstract}

\maketitle

Ordered crystal structures may display long range coherence even if lacking translational symmetry. Materials within this class are quasicrystals, misfit compounds and incommensurate Charge Density Waves (CDWs). In many cases, the lack of a periodic structure gives rise to unconventional transport properties and collective modes \cite{Goldman,Kromer}. Indeed, not much is known about the nature of the electronic states when the Bloch theorem does no longer apply \cite{Rotenberg}. Surely the electronic scattering with the aperiodic potential favor a strong electronic localization. According to the common view, a pseudogap should result as it increases the cohesive energy of the aperiodic structure by a Hume-Rothery mechanism \cite{Fujiwara}. Photoemission experiments on quasicrystals may be consistent with this scenario \cite{Stadnik,Schaub}. Nonetheless, the connection between pseudogap, electronic localization and electron-phonon coupling still has to be settled.

The misfit compound (LaS)$_{1.196}$VS$_2$ is a model system where to study the coupling of an aperiodic potential to the electronic states.  Here, VS$_2$ slabs are stacked with LaS rock-salt planes that are incommensurate in one crystallographic direction \cite{Cario_2005}. At room temperature, this triggers a large modulation of the V-V atomic distance (from 3 and 3.73~\AA~) with a periodicity equal to the lattice mismatch q=1.1~\AA$^{-1}$ \cite{Cario_2005}. The structure of the underlying VS$_2$ slab is similar to the one of other transition metals dichalcogenides, but the modulation amplitude is one order of magnitude larger than in CDWs \cite{Aebi_review,Rossnagel}.
(LaS)$_{1.196}$VS$_2$ may in fact share more properties with quasicrystals. Indeed, a pseudogap was observed by angular integrated photoemission, although it was first attributed to strong correlations \cite{Ino}. Such a pseudogap would explain the lack of metallicity \cite{Cario_1999} observed despite the non integer filling of the band (each LaS unit should transfer one electron to each V, yielding an expected d band filling for V near 2.196). 
Interestingly, this electronic structure seems to host unconventional behaviors. It has been shown that upon the application of moderate electric fields ($\approx$ 50 V/cm), a current switching leads to a volatile increase of conductivity by 6 orders of magnitude \cite{Cario_2006}. Such phenomena could be very interesting for applications \cite{RRAM,Cario2012}.

In this paper, we use Angle and Time Resolved Photoelectron Spectroscopy (ARPES and TRPES) to study the interplay of incommensurate distortion, pseudogap and electron-phonon coupling in (LaS)$_{1.196}$VS$_2$. We observe clearly dispersing bands, qualitatively similar to the ones we calculate for simple undistorted VS$_2$ systems. However, the spectra do not show quasiparticles crossing the Fermi level. Instead, a strongly temperature dependent pseudogap, extending over 100 meV, marks the onset of electronic localization. Two experimental findings support the electron-phonon coupling as main mechanism for the pseudogap formation. First, the temperature dependence of the pseudogap is proportional to the increment of the V-V modulation. Second, the electrons excited by an optical pump release their excess energy to the phonon modes within less than 80 fs. On the same timescale, the electronic pseudogap is partially filled and reaches a metastable state. This timescale is very fast compared to presumably conventional CDW systems (250 fs in 1T-TaS$_2$ \cite{Perfetti_2008} or 500 fs in TbTe$_3$ \cite{Schmitt}). Notice that TRPES is often performed with the idea of separating electronic and structural degrees of freedom, the first ones being typically much faster \cite{Hellmann}. In (LaS)$_{1.196}$VS$_2$, we reach a novel situation, where they are strongly entangled. These results are likely common to other aperiodic structures and quasicrystals.

Single crystals of (LaS)$_{1.196}$VS$_2$ have been synthesized using a self flux method. The ARPES experiments were carried out at the CASSIOPEE beamline of the SOLEIL synchrotron. Photoelectron spectra have been collected with a photon beam of energy 92 eV, linearly polarized in the plane of incidence. The global energy resolution was 10 meV and the angular resolution 0.2$^{\circ}$. Reference band structure calculations for LiVS$_2$ \cite{LiVS2} were performed within the local density approximation, using the Wien2K package \cite{Wien2k}. TRPES experiments were performed with the FemtoARPES setup, using a Ti:Sapphire laser that generates 35 fs pulses centered at 790 nm with repetition rate of 30 kHz.
Part of the beam is employed to generate the fourth harmonic by a cascade of frequency mixing in BBO crystals ($\beta$-BaB$_2$O$_4$) \cite{Faure}. The 197.5 nm probe and the 790 nm pump are focused on the sample with a spot diameter of 100 $\mu$m and 200 $\mu$m, respectively. Their cross-correlation in a BBO crystal has a full width at half maximum (FWHM) of 80 fs. The overall energy resolution of TRPES spectra is limited to 60 meV by the bandwidth of the 197.5 nm beam.

In Fig. 1, we present the room temperature electronic structure measured by conventional ARPES at 92 eV. At first sight, its distribution exhibits 6-fold symmetry, as expected for a triangular lattice. The hexagonal Brillouin Zone (BZ) sketched in red corresponds to a triangular lattice of parameter $a=3.41$~\AA, namely the mean V-V distance observed in the VS$_2$ slab. The highest spectral intensity is found near the zone center $\Gamma$  and the points K. Deep minima are instead found inside large ellipses centered at each M point. 
\begin{figure} \begin{center}
\includegraphics[width=1\columnwidth]{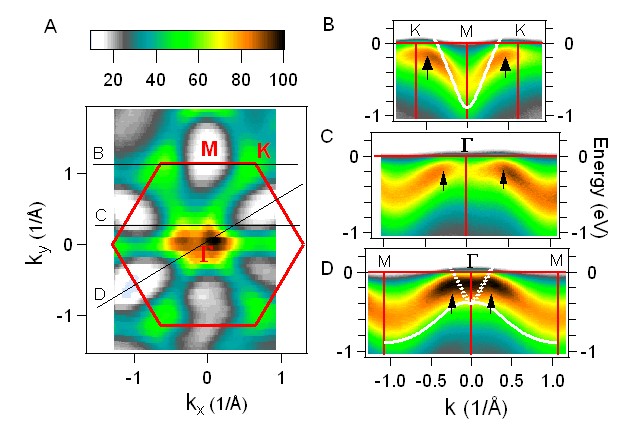}
\caption{A): ARPES spectral weight distribution in $k$-space at 300 K integrated at -40 meV over a 20 meV window. Red lines indicate the BZ of the underlying triangular lattice. B-D): Dispersions along the black lines labelled as B, C and D in A. The white lines in B and D are the dispersions calculated in LiVS$_2$ for $d_{z^2}$ (solid line) and the two other $t_{2g}$ (dotted lines : they are degenerate along $\Gamma$-M and above E$_F$ along K-M-K).} \label{Fig1}
\end{center}
\end{figure}
The band dispersions along different directions are displayed in Fig.~1B-D. In each case, one band is observed with a very clear dispersion, although the peaks are rather broad (about 0.35 eV at half maximum). Therefore, the scattering on the incommensurate modulation does not prevent the dispersion of the electronic states. However, we do not see traces of the incommensurate periodicity, probably because the new supercell zone boundaries are very ill-defined. On the other hand, the bands never cross the Fermi level and we indicate with black arrows the points where they get closest to it.

To fix ideas, we plot on top of the dispersions the band structure calculated in LiVS$_2$, a system with undistorted VS$_2$ slabs and 2 electrons in the V $3d$ orbitals \cite{LiVS2}, close to the filling expected for (LaS)$_{1.196}$VS$_2$. A similar band structure was  observed at lower fillings (3d$^1$) in VS$_2$\cite{MulazziPRB10} or VSe$_2$\cite{StrocovPRL12}. There is a qualitative agreement for the lower band, if the calculation is pushed to higher energy along the $\Gamma$-M direction and down at K. These shifts could be a consequence of the crystal field splitting induced by the distortion in the VS$_6$ octahedra and/or gap openings due to the distortion. We do not clearly observe the upper two t$_{2g}$ bands (dotted lines), although they could still form small electron pockets at $\Gamma$. The lowest band has predominantly $d_{z^2}$ character \cite{Aebi_review}. Its large dispersion of 0.7 eV along $\Gamma$-M is close to that expected in the calculation. This large bandwidth and the large filling factor appear incompatible with the simple picture of a Mott insulating phase in the degenerate $t_{2g}$ manyfold \cite{Ino}.

\begin{figure} \begin{center}
\includegraphics[width=1\columnwidth]{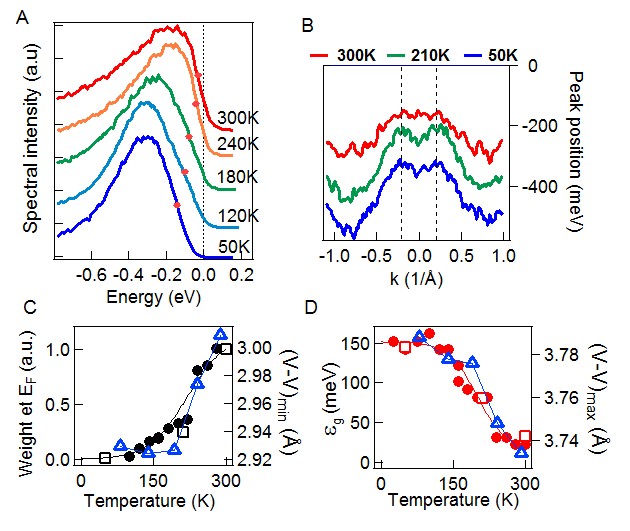}
\caption{A): Photoelectron spectra at $k=0.2$ \AA$^{-1}$ along $\Gamma$-K for the indicated temperatures. B): Dispersion of the band along $\Gamma$-K at 3 different temperatures, obtained by fitting the maximum of the EDC spectra. Dotted lines indicate the location of the pseudogap minimum (arrows in Fig. 1). C): Spectral weight at the Fermi level (left axis, black circles) and minimal value of V-V distance (right axis, blue triangle). D): Leading edge pseudogap $\varepsilon_g$ (left axis, red circles in A) and maximal value of V-V distance (right axis, blue triangle). Filled circles and open squares correspond to 2 different samples.} \label{Fig2}
\end{center}
\end{figure}

In Fig. 2A, we show spectra at $k=0.2$~\AA$^{-1}$, corresponding to the pseudogap minimum (dotted line in Fig. 2B). The lack of a clear Fermi crossing is the major deviation with respect to the undistorted VS$_2$ band structure. It leads to a strong reduction of spectral weight at the Fermi level and we will refer to this situation as pseudogap.
When the temperature is lowered, the main peak shifts to higher binding energy by about 100 meV. More precisely, the whole band shifts quite uniformly, as detailed in Fig. 2B for the $\Gamma$-K direction. This is different from the opening of a gap expected in the weak coupling limit of a CDW. As the compound becomes quite insulating in the process \cite{Cario_2006}, we checked at each temperature that there was no shift due to charging effects \cite{Charging}. In Fig. 2C, we plot the vanishing of the spectral density at E$_F$ as a function of temperature and in Fig. 2D the shift of the leading edge $\varepsilon_g$ (red points in Fig. 2A). There is no unique definition of a pseudogap, but similar temperature dependences of the shift were obtained from the peak position or the center of gravity. These two quantities scale with the minimal (Fig. 2C) and maximal (Fig. 2D) values of the V-V distances (blue triangles) as a function of temperature. Indeed, the distortion amplitude increases by 0.05~\AA~between 300 K and 100 K, although there is no structural phase transition \cite{Vinh_Preprint}. This establishes strongly a link between the aperiodic V-V modulation and the pseudogap formation.




\begin{figure} \begin{center}
\includegraphics[width=1\columnwidth]{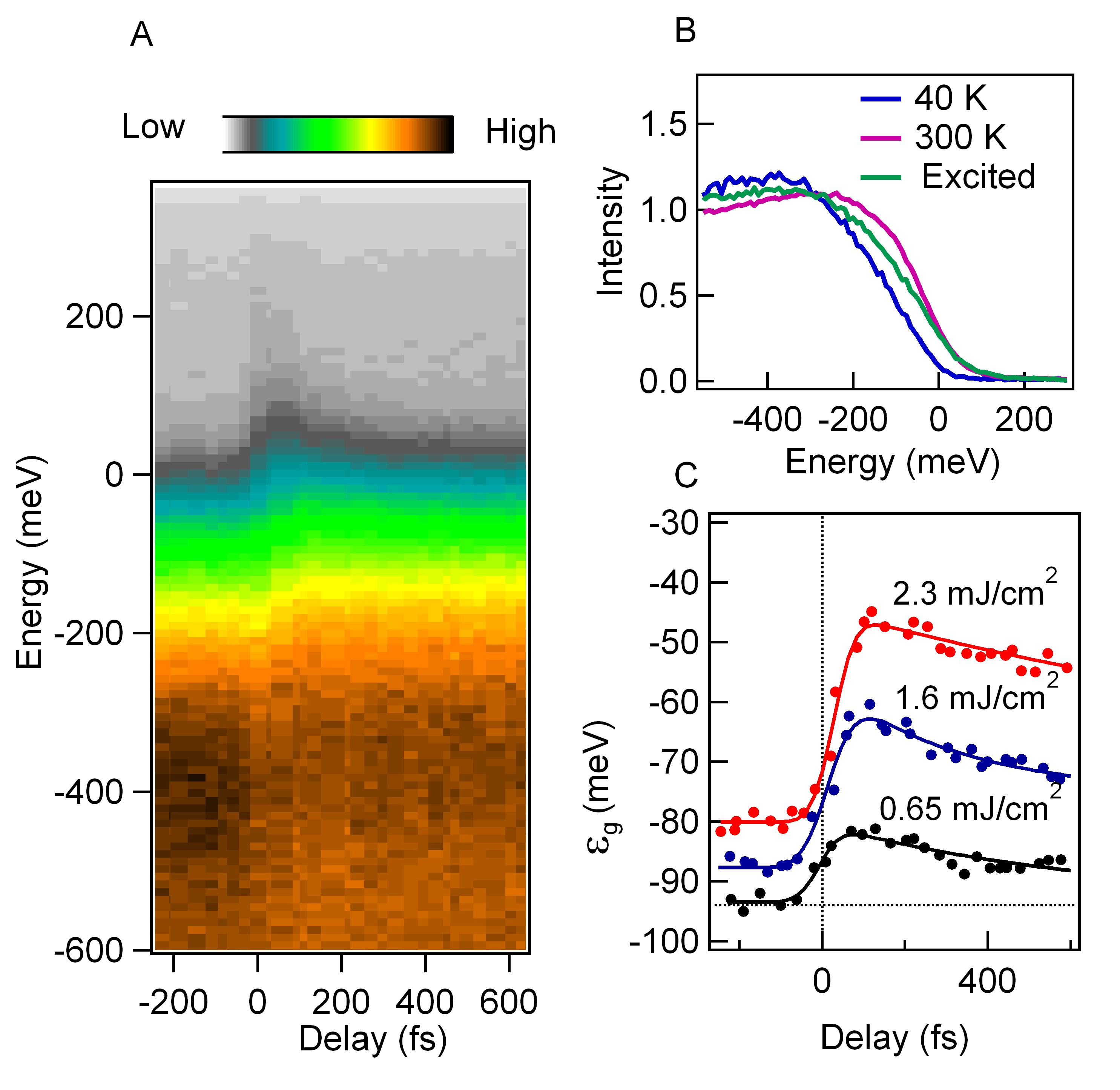}
\caption{A): Intensity map of photoelectrons emitted normal to the sample surface as a function of pump probe delay, for an incident fluence of 2.3 mJ/cm$^2$. B): Photoelectron spectra acquired in equilibrium at 40 K (blue curve) and 300 K (violet) compared to the spectrum acquired at the base temperature of 40 K and 100 fs after the arrival of the pump beam (green curve). C): Sudden reduction and subsequent recovery of the electronic pseudogap for several fluences of the incident pump beam.} \label{Fig3}
\end{center}
\end{figure}

To better understand the relevance of electron-phonon coupling on the charge gap we investigate next the temporal evolution of the electronic states after photoexcitation by an intense laser pulse. Figure \ref{Fig3}A displays a photoelectron intensity map acquired at 40 K as a function of pump-probe delay. The photoelectrons generated by the 6.3 eV probe beam are detected in a small angular window centered around normal emission. Pump pulses with incident fluence of 2.3 mJ/cm$^2$ induce an excitation density of roughly 0.065 electrons per (LaS)$_{1.196}$VS$_2$ unit. The photoexcitation clearly increases the signal between -250 meV and the Fermi level. In this spectral range, the partial filling of the electronic pseudogap overwhelms the photoinduced change of the electronic occupation factor. Nonetheless, the non-equilibrium distribution of the electrons is still visible in the counts above the Fermi level.

We show in Fig. \ref{Fig3}B, the photoelectron spectra, extracted from the intensity map of Fig. \ref{Fig3}A, 100 fs after the arrival of the pump beam (green curve). As a term of comparison we also plot the photoelectron spectra in thermal equilibrium at 40 K (blue curve) and 300 K (violet curve). 
When the sample is at 40 K and the pump beam is off, the value of the leading edge is $\varepsilon_g=-95$ meV \cite{Leading Edge}. 

Upon the arrival of the pump pulse, the leading edge suddenly approaches the Fermi level. This reduction of the electronic pseudogap is proportional to the incident fluence and therefore to the internal energy deposited into the system. We fit the early dynamics of $\varepsilon_g$ by the convolution of $\theta(t-t_0)e^{-t/\tau}$ with a gaussian function. The width of the gaussian function is $100$ fs and turns out to be only $20$ fs larger than the cross correlation between pump and probe pulse. The parameter $t_0$ is 30 fs larger than zero only for the curve obtained with incident fluence of 2.3 mJ/cm$^2$. We conclude that the pseudogap filling arises on a very short timescale, comparable to or shorter than the temporal resolution of our experiment.

\begin{figure} \begin{center}
\includegraphics[width=1\columnwidth]{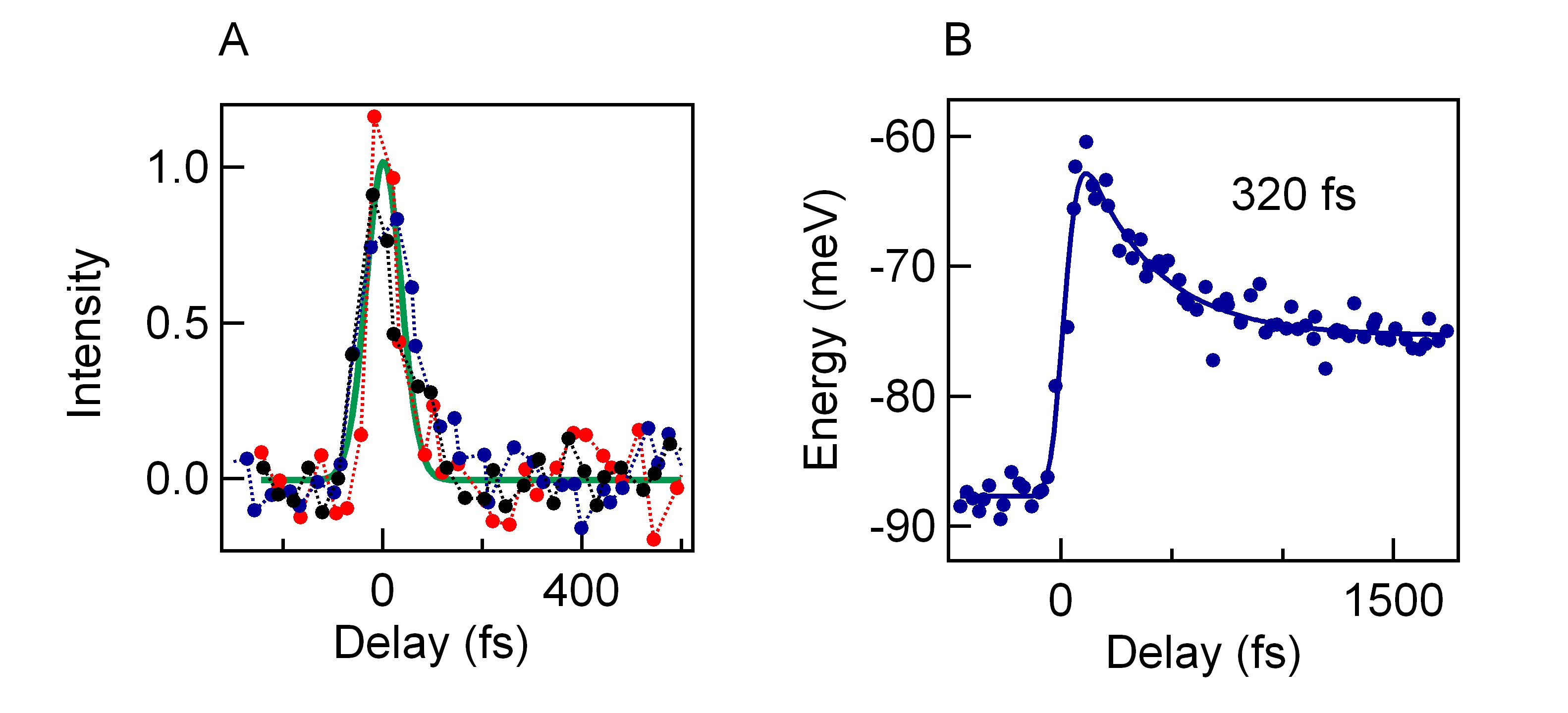}
\caption{A): Temporal evolution of the photoelectron current acquired at energy $\varepsilon=220$ meV above the Fermi level. The red, blue and black dots stands for measurements with incident pump fluence of 2.3 mJ/cm$^2$, 1.6 mJ/cm$^2$ and 0.6 mJ/cm$^2$, respectively. The green line superimposed to the data is the measured cross correlation between our pump and probe pulse. B): Evolution of the electronic pseudogap as a function of pump probe delay. } \label{Fig4}
\end{center}
\end{figure}

The depopulation of the electronic states above the Fermi level confirms the occurrence of an abrupt energy flow in lattice modes. We show in Fig. \ref{Fig4}A the temporal evolution of the renormalized photoelectron signal acquired for $\varepsilon = 220$ meV. The transient electronic occupation displays a gaussian shape which is comparable to the cross-correlation between pump and probe pulse. We deduce that heat is transferred from electrons to the phonons on a timescale shorter than $\Delta t$=80 fs, which is very fast compared to known CDW systems \cite{Perfetti_2008,Schmitt}.

The temporal evolution of Fig. 4B indicates that electron-phonon system at 100 fs is still far from equilibrium conditions. Only the strongly coupled modes are efficiently populated, whereas the others can be considered as still frozen.
Such selective heating has already been reported in other systems \cite{Perfetti_2007,Kampfrath}. The warming up process is followed by the reduction of free energy via the anharmonic scattering between highly excited modes and colder ones. In agreement with this mechanism, Fig. \ref{Fig4}B shows that the electronic pseudogap partially recovers with time constant of 320 fs.

Next, we discuss how non-equilibrium phonons affect the electronic structure.
Qualitatively speaking, two complementary effects lead to the collapse of the electronic pseudogap. First, when photoexcited electrons transfer their energy to some phonon modes, the mean standard deviation of the atomic displacements becomes much larger than in the unpumped system. It follows the emergence of intragap electronic excitations that are likely localized on a rather short lengthscale. Second, the photoinduced change of the free energy suddenly switches on a force on the atomic lattice. 
Therefore, the resulting atomic displacement should oscillate at the characteristic frequency of the related mode \cite{Fritz}. In this case, the coupling between the lattice and the electronic system would modulate the electronic pseudogap \cite{Perfetti_2008, Schmitt, Hellmann}.

Concerning (LaS)$_{1.196}$VS$_2$, both mechanisms are in principle possible. However, the photoelectron map in Fig. 3 does not display any periodic modulation of the spectral density. It is not excluded that coherent phonons of small amplitude lie below the noise threshold of our measurement. Moreover, the 80 fs temporal duration of our probe pulse would hinder the detection of oscillations with frequency higher than 5 THz (20meV). The phonon cut-off is not known for (LaS)$_{1.196}$VS$_2$, but it is as high as 80meV in 1T-VSe$_2$ \cite{Kamarchuk}, so that we may average in time the effects due to fast modes. Nonetheless, the different spectral shapes in the photoexcited and high temperature cases (see. Fig. 3B) also suggest that photoexcitation fills the electronic pseudogap rather than closes it and that this process is dominated by incoherent lattice motion instead of a coherent one. 





In conclusion, (LaS)$_{1.196}$VS$_2$ is an original example of an insulator with large sensitivity to external parameters such as electric field \cite{Cario_2006}, temperature or laser pulses. Our measurements disclose the interplay between aperiodicity, pseudogap and strong electron phonon coupling. Despite the incommensurate potential, the electronic structure is rather simple and follows calculations based on a triangular system. The filling and bandwidth of the $d_{z^2}$ are not compatible with a Mott scenario of the insulating phase. Instead, our measurements indicate that strong electron-phonon coupling generates the pseudogap at the Fermi level. The  temperature dependence of the pseudogap nearly follows the increase of V-V modulation measured by X-ray diffraction. Upon photoexcitation the pseudogap is filled by localized electronic state in less than 100 fs and reaches a metastable phase. This shows that, even when the structural degree of freedom obviously plays a major role, a very fast dynamic can be observed. Our measurement establishes that the electronic energy is transferred to phonon modes in less than 80 fs. This reveals the essential role of strong electron-phonon coupling in electronic localization in quasiperiodic structures and quasicrystals.

We thank N. Vast for useful discussions. The FemtoARPES project was financially supported by the RTRA Triangle
de la Physique, the ANR program Chaires d'Excellence (Nr. ANR-08-CEXCEC8-011-01).


\begin{thebibliography}{99}
\bibitem{Charging}
We ruled out charging effects by reducing the photon flux by a factor 3 without inducing any shift down to 30 K.
\bibitem{Leading Edge}
The pseudogap observed at 92 eV is slightly smaller, which might be due to the different integration window and different cross section at 6 eV. 
The presence of the pump beam slightly reduces the magnitude of  $\varepsilon_g$ at negative pump-probe delay (see Fig. \ref{Fig3}C), but this does not exceed 15 meV. We find that a laser repetition rate of 30 kHz is the optimal compromise between the accumulative heating and good statistics.
\end{thebibliography}
\end{document}